\documentclass{article}
\usepackage{spconf,amsmath,graphicx}
\usepackage{amsmath,multirow,array,enumitem,adjustbox}
\usepackage{subcaption,booktabs}

\usepackage{hyperref}
\hypersetup{colorlinks=true,citecolor=green}

\usepackage{cleveref}
\usepackage{threeparttable}  %

\usepackage{caption}
\makeatletter
\renewcommand\section{\@startsection {section}{1}{\z@}%
                                   {-2.5ex \@plus -1ex \@minus -.2ex}%
                                   {1.75ex \@plus.2ex}%
                                   {\normalfont\Large\bfseries}}
\renewcommand\subsection{\@startsection{subsection}{2}{\z@}%
                                     {-1.5ex\@plus -1ex \@minus -.2ex}%
                                     {1.ex \@plus .2ex}%
                                     {\normalfont\large\bfseries}}
\renewcommand\subsubsection{\@startsection{subsubsection}{3}{\z@}%
                                     {-3.25ex\@plus -1ex \@minus -.2ex}%
                                     {1.5ex \@plus .2ex}%
                                     {\normalfont\normalsize\bfseries}}
\renewcommand\paragraph{\@startsection{paragraph}{4}{\z@}%
                                    {3.25ex \@plus1ex \@minus.2ex}%
                                    {-1em}%
                                    {\normalfont\normalsize\bfseries}}
\renewcommand\subparagraph{\@startsection{subparagraph}{5}{\parindent}%
                                    {3.25ex \@plus1ex \@minus .2ex}%
                                    {-1em}%
                                    {\normalfont\normalsize\bfseries}}
\makeatother

\title{Self-Supervised Learning Based Domain Adaptation for Robust Speaker Verification}
\name{Zhengyang Chen, Shuai Wang, Yanmin Qian\textsuperscript{\dag} \thanks{\textsuperscript{\dag}Yanmin Qian is the corresponding author}}
\address{MoE Key Lab of Artificial Intelligence, AI Institute \\
    SpeechLab, Department of Computer Science and Engineering \\
    Shanghai Jiao Tong University, Shanghai, China\\
    {\small \{zhengyang.chen, feixiang121976, yanminqian\}@sjtu.edu.cn}}
\begin{document}
\ninept
\maketitle
\begin{abstract}
Large performance degradation is often observed for speaker verification systems when applied to a new domain dataset. Given an unlabeled target-domain dataset, unsupervised domain adaptation (UDA) methods, which usually leverage adversarial training strategies, are commonly used to bridge the performance gap caused by the domain mismatch. However, such adversarial training strategy only uses the distribution information of target domain data and can not ensure the performance improvement on the target domain. In this paper, we incorporate self-supervised learning strategy to the unsupervised domain adaptation system and proposed a self-supervised learning based domain adaptation approach (SSDA). Compared to the traditional UDA method, the new SSDA training strategy can fully leverage the potential label information from target domain and adapt the speaker discrimination ability from source domain simultaneously. We evaluated the proposed approach on the VoxCeleb (labeled source domain) and CnCeleb (unlabeled target domain) datasets, and the best SSDA system obtains 10.2\% Equal Error Rate (EER) on the CnCeleb dataset without using any speaker labels on CnCeleb, which also can achieve the state-of-the-art results on this corpus. 

\end{abstract}
\section{Introduction}
\label{sec:intro}

Speaker verification aims to verify  a person's identity given his or her voice. In recent years, the thriving of deep neural network (DNN) has led to great success of speaker verification systems~\cite{arc_1}. To improve the performance and robustness of speaker verification systems, researchers have designed different network backbones ~\cite{arc_2,xvector,r-vector}, different pooling functions~\cite{pooling1,pooling2,pooling3} and loss functions \cite{loss_1,loss_2,loss_4,huang_asoftmax,loss_5}.

However, DNN based speaker verification systems usually require a large amount of well-labeled data for training, which is not available in most cases. On the other hand, a well-trained speaker verification system suffers from severe performance degradation when adapted to another dataset from a different domain.  Thus, it is necessary to develop a method to fast adapt an existing model trained on well-labeled source domain data to a new target domain dataset where no speaker label is available. Such a task is considered as unsupervised domain adaptation (UDA) \cite{uda1,uda2}.

Researchers have proposed different UDA methods to tackle this problem. The most common practice is using adversarial training strategy \cite{but,adv_spk1,adv_spk2, adv_spk3} to minimize the distribution mismatch of the learned embeddings from different domains, which is expected to maintain the speaker discrimination ability learned from the well-labeled source domain data to the target domain. Other researchers have tried to use clustering methods \cite{uda_cluster} to estimate pseudo-labels for unlabeled target domain data and then do supervised training using the estimated labels. However, in adversarial training-based UDA methods, only the data distribution information from a different domain is used, and too aggressively matching the embedding distribution from different domains may hurt the speaker discrimination ability to some extent \cite{adv_spk3}. Besides, in clustering-based UDA methods, it is hard to determine the speaker number when doing clustering and the estimated label may not be accurate.

To fully leverage the available information from the target domain dataset, we proposed a self-supervised learning based domain adaptation method (SSDA) for unsupervised domain adaptation. 
According to the continuity of speech, there is usually only one person speaking in short time duration. Here, we assume that each utterance only contains one speaker, which is true for many datasets\footnote{Even for those conversational speeches, we still can easily prepare the data by using techniques such as speaker diarization}. We can then sample positive pairs from the same utterance and sample negative pairs from different utterances to do contrastive learning \cite{contrastive_2} without reaching the speaker labels. Although purely self-supervised learning could be done with an unlabeled target domain dataset, the performance is usually not guaranteed. 
In this paper, we apply this self-supervised learning strategy to adapt a well-trained source speaker embedding extractor to a target domain dataset, simultaneously taking advantage of the rich information in both the source and target datasets. Experiments are carried out on the VoxCeleb and CnCeleb dataset.
Our best SSDA system achieves 10.2\% EER on the CnCeleb dataset even without using any speaker label and exceeds the current state-of-the-art result. Besides, when speaker labels from CnCeleb are used and source and target domain data are jointly trained, the system makes further improvement and achieves 8.86\% EER on the CnCeleb evaluation set.

\section{Method}
\label{sec:method}

This section will first give a brief introduction to the contrastive learning strategy, which is the foundation of self-supervised training. Then, four objective functions for constrastive learning will be introduced, and we will implement and compare their effects for our SSDA experiments in the following sections. Finally, we will present the self-supervised adaptation (SSDA) algorithm, together with the system configuration and training strategies.

\subsection{Contrastive Learning}
Contrastive learning aims to maximize the similarity of positive pairs and minimize the similarity of negative pairs. Recently, many researchers have been studying to construct contrastive pairs based on an unlabeled dataset to do self-supervised training. For computer visual representation learning, researchers have used augmentation methods to construct positive and negative pairs for contrastive learning  \cite{contrastive_1,contrastive_2} and any label is not needed here. Because of the continuity of speech, there is always only one person speaking in short time duration. Thus, for speaker representation learning, there is no need to do data augmentation for positive pair sampling and we can randomly sample two segments from one utterance and consider them as a positive pair. Similarly, we can consider two segments sampled from different utterances as a negative pair.

\subsection{Contrastive Learning Objectives for Self-Supervised Training}
\label{ssec:contrastive_obj}
For contrastive learning objectives, following the implementation in \cite{metric_loss}, we randomly sample M segments from each of N utterances, whose embeddings are $x_{j,i}$ where $1 \leq j \leq N \text{ and } 1 \leq i \leq M$. The segments sampled from the same utterance are considered from the same speaker and segments from different utterance are considered from different speaker.

\subsubsection{Contrastive Loss}
The contrastive loss aims to maximize the distance of negative pairs and minimize the positive pairs' distance in a mini-batch. As shown in equation \ref{eq:loss_contrastive}, positive pairs are sampled from the same utterance and negative pairs are sampled from different utterances within the mini-batch based on a hard negative mining strategy, which requires $M=2$ in this condition. The margin $m$ is set to $4.0$ here.

\begin{equation}
\label{eq:loss_contrastive}
\begin{split}
\mathcal{L}_{C} &= \frac{1}{N} \sum_{j=1}^{N}\left\|\mathbf{x}_{j, 1}-\mathbf{x}_{j, 2}\right\|_{2}^{2} \\
&+  \frac{1}{N} \sum_{j=1}^{N} \max(0, m - \left\|\mathbf{x}_{j, 1}-\mathbf{x}_{k \neq j, 2}\right\|_{2}^{2})
\end{split}
\end{equation}

\subsubsection{Triplet Loss}
Triplet loss minimizes the L2 distance between an anchor and a positive, and maximizes the distance between an anchor and a negative. As shown in equation \ref{eq:loss_triplet}, $x_k$ is also sampled using a hard negative mining strategy as the last section and $M$ is set to 2. The margin $m$ is set to $4.0$ here.

\begin{equation}
\label{eq:loss_triplet}
\begin{split}
L_{\mathrm{T}}=\frac{1}{N} \sum_{j=1}^{N} \max \left(0,\left\|\mathbf{x}_{j, 1}-\mathbf{x}_{j, 2}\right\|_{2}^{2}-\left\|\mathbf{x}_{j, 1}-\mathbf{x}_{k \neq j, 2}\right\|_{2}^{2}+m\right)
\end{split}
\end{equation}

\subsubsection{Angular Prototypical Loss}
Prototypical loss can also be considered as a kind of contrastive learning objectives, where the pair is constructed between a centriod and a query. For prototypical loss (ProtoLoss), each mini-batch contains a support set $S$ and a query set $Q$. Same as the implementation in \cite{metric_loss}, the $M$-th segment from each utterance is considered as query. Then the prototype (centroid) is defined as:

\begin{equation}
\label{eq:loss_proto_sub1}
\begin{split}
\mathbf{c}_{j}=\frac{1}{M-1} \sum_{m=1}^{M-1} \mathbf{x}_{j, m}
\end{split}
\end{equation}

Here, we use the angular version prototypical loss and cosine score is used to measure the similarity. Different from the original format in \cite{metric_loss}, we add a temperature hyper-parameter $\tau$ to the cosine similarity which controls the concentration level of the distribution following \cite{temperature} and the score is defined as:
\begin{equation}
\label{eq:loss_proto_sub2}
\begin{split}
\mathbf{S}_{j, k}=\tau \cdot \cos \left(\mathbf{x}_{j, M}, \mathbf{c}_{k}\right)
\end{split}
\end{equation}

The angular prototypical loss is calculated by a softmax function, in which each query is classified against N centroids:
\begin{equation}
\label{eq:loss_proto}
\begin{split}
L_{\mathrm{P}}=-\frac{1}{N} \sum_{j=1}^{N} \log \frac{e^{\mathbf{S}_{j, j}}}{\sum_{k=1}^{N} e^{\mathbf{S}_{j, k}}}
\end{split}
\end{equation}

\subsubsection{Generalised End to End Loss (GE2E)}
Different from prototypical loss, in GE2E, each segment in a mini-batch is used to calculate the centroids: 

\begin{equation}
\label{eq:loss_ge2e_sub1}
\begin{split}
\mathbf{c}_{j}=\frac{1}{M} \sum_{m=1}^{M} \mathbf{x}_{j, m}
\end{split}
\end{equation}

\begin{equation}
\label{eq:loss_ge2e_sub2}
\begin{split}
\mathbf{c}_{j}^{(-i)}=\frac{1}{M-1} \sum_{m=1 \atop m \neq i}^{M} \mathbf{x}_{j, m}
\end{split}
\end{equation}

Here, each segment is also used as a query. When query and the centroid are from the same utterance, the query itself is excluded when calculating the centroid. We also add a temperature hyper-parameter $\tau$ to the cosine similarity:
\begin{equation}
\label{eq:loss_ge2e_sub3}
\begin{split}
\mathbf{S}_{j, i, k}=\left\{\begin{array}{ll}
\tau \cdot \cos \left(\mathbf{x}_{j, i}, \mathbf{c}_{j}^{(-i)}\right) & \text { if } \quad k=j \\
\tau \cdot \cos \left(\mathbf{x}_{j, i}, \mathbf{c}_{k}\right) & \text { otherwise }
\end{array}\right.
\end{split}
\end{equation}

The GE2E loss is defined as:

\begin{equation}
\label{eq:loss_ge2e}
\begin{split}
L_{\mathrm{G}}=-\frac{1}{N \cdot M} \sum_{j, i} \log \frac{e^{\mathbf{S}_{j, i, j}}}{\sum_{k=1}^{N} e^{\mathbf{S}_{j, i, k}}}
\end{split}
\end{equation}

\subsection{Domain Adaptation with Self-Supervised Learning}
\label{ssec:intro_sda_both}

\subsubsection{Self-Supervised adaptation training}
\label{sssec:intro_sda}
In this section, we first introduce a simple self-supervised learning based domain adaptation (\textbf{SSDA}) training strategy. To adapt the model from the source domain, we first initialized the embedding extractor by the well-trained model from the source domain. Then, we do self-supervised training on the target domain data based on the contrastive learning objectives in section \ref{ssec:contrastive_obj}.

\begin{figure}[!h]
\centering
\includegraphics[width=0.45\textwidth]{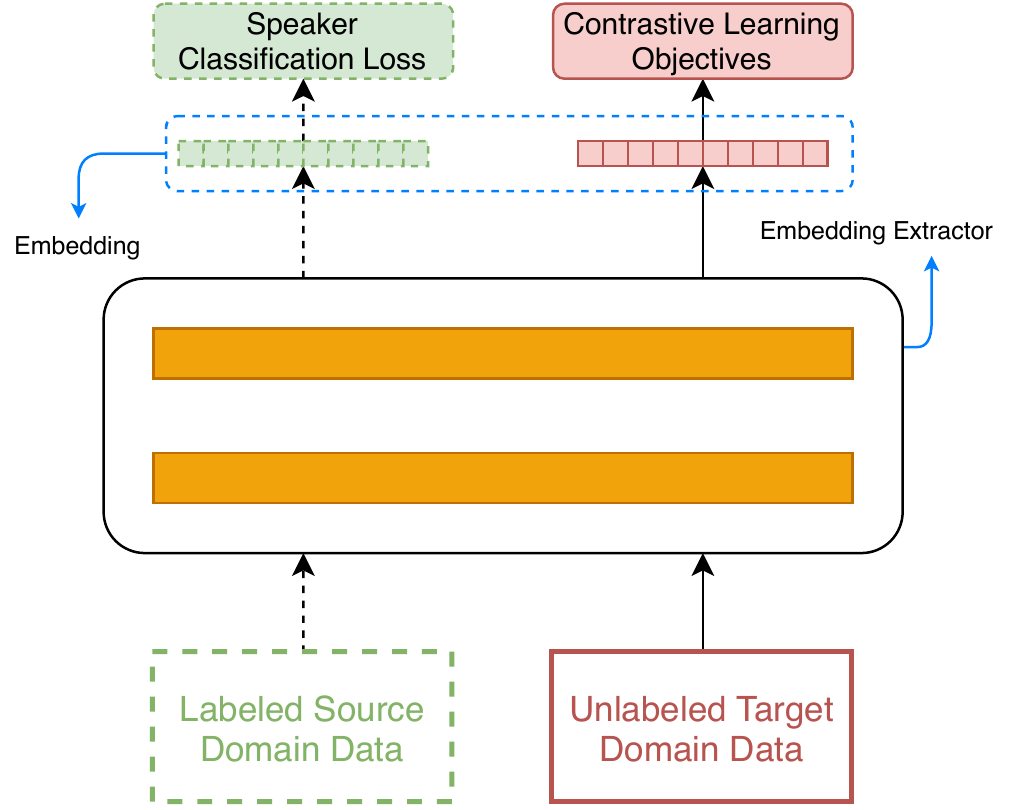}
\caption{Self-Supervised learning based domain adaptation.}
\label{fig:sda}
\end{figure}

As shown in the solid line part of Figure \ref{fig:sda}, the embedding extractor is only supervised by the contrastive learning loss $L_{cl}$ during this adaptation.

\subsubsection{Self-Supervised adaptation with joint training}
\label{sssec:intro_sda_joint}
Simply using the self-supervised learning strategy may cause the overfitting on the target domain data and the embedding extractor may lose the strong speaker discrimination ability, which benefits from source domain pre-training. Here, we proposed a joint training strategy for the SSDA, and it is named as \textbf{SSDA-Joint}.

As shown in Figure \ref{fig:sda}, the labeled source domain data and unlabeled target domain data are fed into the embedding extractor simultaneously during the training stage. The output source domain embeddings are supervised by the speaker classification loss and target domain embeddings are supervised by the contrastive learning loss. These two losses are jointly optimized and the total loss is defined as:

\begin{equation}
\label{eq:loss_total}
\begin{split}
L_{total}= L_{cla} + \lambda L_{cl}
\end{split}
\end{equation}
where $L_{cla}$ is the speaker classification loss and $\lambda$ is a hyper-parameter for weighted summation.

\section{Experimental Setup}
\label{sec:exp_setup}

\subsection{Dataset}
\label{ssec:dataset_setup}
To perform domain adaptation, we use two datasets from different languages in our experiments and the detailed information of training parts for these two datasets is shown in Table \ref{table:train_data}.
In our experiment, the data from VoxCeleb1 \& 2 \cite{vox} is considered as source domain data. The source domain data is downloaded from Youtube and most of the speech data is in English. Besides, the data from CnCeleb \cite{CnCeleb} is considered as target domain data. The target domain data is downloaded from Bilibili and the speech data is in Chinese. 

The 5994 speakers from the VoxCeleb2 dev set are used as source domain training data and the whole VoxCeleb1 is used as source domain evaluation set. For target domain data, following the official split, 800 speakers are used for training and 200 speakers are used for evaluation. It should be noted that we don't use any speaker label when doing target domain self-supervised learning.

We use the official trial list for CnCeleb to evaluate our system. This trial list contains 18,024 target pairs and 3,586,776 nontarget pairs. Besides, when we evaluate our system on the source domain dataset, we consider the whole Voxceleb1 set as the evaluation dataset and all three official trial lists Vox1-O, Vox1-E and Vox1-H are used for evaluation.

\begin{table}[h!]
\footnotesize
\centering
\caption{\textbf{Training data information.} The CnCeleb training utterance number is the statistic after preprocessing which is described section \ref{ssec:data_process_setup}.}
\begin{adjustbox}{width=.4\textwidth,center}
\begin{tabular}{c|c|c}
    \toprule 
    
     & Source Domain & Target Domain \\
    \hline
    Data & VoxCeleb & CnCeleb \\
    
    Language & Mostly English  & Chinese  \\
    \# of Spk & 5,994 & 800 \\
    \# of Utt & 1,021,161 & 53,288 \\
    \# of Hour & 2,207  & 148 \\
    
    \bottomrule
\end{tabular}
\label{table:train_data}
\end{adjustbox}
\end{table}

\subsection{Data Preprocessing}
\label{ssec:data_process_setup}

For CnCeleb training data, we first combine the short utterances to make them longer than 5 seconds because there are too many very short utterances. After processing, there are 53288 utterances left. For all the audio data, 40-dimensional Fbank features are extracted using Kaldi toolkit~\cite{povey2011kaldi} with a 25ms window and 10ms frame shift, and silence is removed using an energy-based voice activity detector. Then we do the cepstral mean on the Fbank features with a sliding-window size of 300 frames. Similar to the Kaldi VoxCeleb recipe, we also discard all the utterances of less than 400 frames.

\subsection{System Configuration}
\label{ssec:system_setup}
Resnet based r-vector \cite{r-vector} is used as the embedding extractor in our experiment and the embedding dimension is set to 256. In our experiments, we find it is better to set $M=2$ for ProtoLoss and GE2E and we set the temperature hyper-parameter $\tau$ to 32. 
When additive angular margin (AAM) loss \cite{loss_2} is used in our experiment, margin is set to 0.2 for source domain data and 0.15 for target domain data. For simplicity, we set $\lambda=1$ in equation \ref{eq:loss_total} for SSDA-Joint training loss. Besides, the pre-trained model from source domain is also trained using AAM loss with the same configuration. For all the experiments in this paper, we use cosine similarity to score the trials.

\section{Results}
\label{sec:res}

\subsection{Comparison between Supervised and Self-Supervised Baselines.}

\begin{table}[h!]
\footnotesize
\centering
\caption{\textbf{Comparison between supervised and self-supervised results on the CnCeleb.} }
\begin{adjustbox}{width=.45\textwidth,center}
\begin{threeparttable}
\begin{tabular}{cccc}
    \toprule 
    Train Data & Train Mode & Loss & EER (\%) \\
    \hline
    VoxCeleb$^*$ & \multirow{3}{*}{Supervised} & AAM & 12.11  \\
    CnCeleb &  & Softmax & 14.16  \\
    CnCeleb &  & AAM & 13.43  \\
    
    \hline
    \multirow{4}{*}{CnCeleb} & \multirow{4}{*}{Self-Supervised} & Contrastive &  23.00 \\
    &  & Triplet &  20.40 \\
    &  & ProtoLoss &  18.97 \\
    &  & GE2E & 18.76 \\

    \bottomrule
\end{tabular}
\begin{tablenotes}\footnotesize
\item $^*$This line is considered as the baseline system for our experiment. Here, only labeled VoxCeleb data is used to train the embedding extractor and the trained model is directly tested on the CnCeleb evaluation set.
\end{tablenotes}
\end{threeparttable}
\label{table:res_baseline}
\end{adjustbox}
\end{table}

In this section, we first evaluate the performance of self-supervised training strategy with different contrastive objectives and compare it with supervised training. Results are shown in Table \ref{table:res_baseline}. Obviously, with CnCeleb speaker labels available, the supervised training results are better than all the self-supervised training ones. Encouragingly, we find the performance self-supervised training is not that bad and the performance of the best self-supervised system is close to the Cnceleb supervised training result. Besides, the model trained on the VoxCeleb data in supervised mode performs better than the Cnceleb ones, mainly because the data amount of source domain is much larger than the target domain as shown in Table \ref{table:train_data}. Here, we consider the result of the model trained on Voxceleb in supervised training mode as the baseline for the following experiments and such result can be considered as a simple domain adaptation from source domain to target domain.

\subsection{Self-Supervised Learning based Domain Adaptation.}
The proposed SSDA and SSDA-Joint training strategy introduced in section \ref{ssec:intro_sda_both} are evaluated in this section and the results are shown in Table \ref{table:res_sda}. 

\begin{table}[h!]
\footnotesize
\centering
\caption{\textbf{Self-Supervised learning based domain adaptation results.}}
\begin{adjustbox}{width=.45\textwidth,center}
\begin{threeparttable}
\begin{tabular}{cccc}
    \toprule 
    
    Train Data & Training Mode & Target Loss & EER (\%) \\
    \hline
    VoxCeleb & Supervised & - & 12.11  \\
    
    \hline
    
    \multirow{8}{*}{VoxCeleb+CnCeleb} & \multirow{4}{*}{SSDA} & Contrastive & 20.72 \\
    &  & Triplet & 13.66 \\
    &  & ProtoLoss & 13.72 \\
    &  & GE2E & \textbf{13.34} \\
    \cline{2-4}
     & \multirow{4}{*}{SSDA-Joint} & Contrastive  & 20.48  \\
    &  & Triplet  & 11.27 \\
    &  & ProtoLoss  & \textbf{10.20} \\
    &  & GE2E  & 10.24 \\
    
    \bottomrule
\end{tabular}

\end{threeparttable}
\label{table:res_sda}
\end{adjustbox}
\end{table}

\subsubsection{SSDA training result}
From the upper part of Table \ref{table:res_sda}, we find that the performance of our proposed SSDA training strategy exceeds the simple target domain self-supervised learning by a substantial margin, which confirms that a well-pretrained model on the large scale source domain data is important. However, the SSDA training strategy still perform worse than the strong baseline system. As we assumed in section \ref{sssec:intro_sda}, simply applying self-supervised training on the target domain data may cause overfitting and the embedding extractor may lose the strong speaker discrimination ability adapted from source domain after target domain training.

\subsubsection{SSDA-Joint training result}
Plus the joint training with source domain data, SSDA-Joint training strategy further improves target domain performance compared to SSDA strategy. Notably, the SSDA-Joint training strategy with most of the contrastive learning losses performs better than the strong baseline system. Besides, the ProtoLoss performs the best in this condition and achieves $15.7\%$ relative improvement in terms of EER compared to baseline. To the best of our knowledge, such a result exceeds the current state-of-the-art result on the CnCeleb evaluation set even without using any target domain speaker label.

\subsubsection{Evaluation on the source domain dataset}
Besides, we evaluated our best SSDA-Joint (ProtoLoss) model on the source domain evaluation set and results are shown in Table \ref{table:res_source_domain}.
Surprisingly, we find that the adaptation model trained with the SSDA-Joint training strategy performs even better than the original source domain well-trained model. The possible explanation is that joint training with target domain data may have a regularization effect on the original source domain task and much more available training data in the training process can make the embedding extractor more robust.

\begin{table}[h!]
\footnotesize
\centering
\caption{\textbf{Performance of SSDA-Joint model on source domain evaluation set.} The SSDA-Joint model is evaluated on the VoxCeleb1 dataset and three official trials for VoxCeleb1 are used. }
\begin{adjustbox}{width=.4\textwidth,center}
\begin{tabular}{cccc}
    \toprule 
    \multirow{2}{*}{Model} &
    \multicolumn{3}{c}{EER (\%)}\\
    \cline{2-4}
    & Vox-O & Vox-E & Vox-H  \\
    \hline
    Source Train & 1.77 & 1.77 & 3.07 \\
    After Adaptation & \textbf{1.56} & \textbf{1.73} & \textbf{3.00} \\
    \bottomrule
\end{tabular}
\label{table:res_source_domain}
\end{adjustbox}
\end{table}

\subsubsection{Comparison with fully supervised-joint training results}

To further explore the gap between the unsupervised domain adaptation results and the fully supervised joint training results, we list the relevant results in the Table \ref{table:res_supervise}. The supervised-joint training result can be considered as a upper-bound performance on the CnCeleb in our experiment. Notably, we find the gap between the fully supervised-joint training method and the SSDA-Joint training method is small, which means that we can save a large amount of labeling cost with minor performance degradation on the target domain by using our proposed adaptation method.

\begin{table}[!h]
\footnotesize
\centering
\caption{\textbf{Comparison with supervised-joint training results}. For Supervised-Joint mode, the target domain contrastive learning loss is replaced by the supervised speaker classification loss. The source and target domain classification losses are also weighted summed as equation \ref{eq:loss_total}.}
\begin{adjustbox}{width=.48\textwidth,center}
\begin{tabular}{cccc}
    \toprule 

    Train Data & Train Mode & Target Loss & EER (\%) \\
    
    \hline
    \multirow{4}{*}{VoxCeleb+CnCeleb} & \multirow{2}{*}{SSDA-Joint} & GE2E       & 10.24  \\
     &  & ProtoLoss & 10.20 \\
    \cline{2-4}
     & \multirow{2}{*}{Supervised-Joint} & Softmax & 8.860 \\
     &  & AAM & 9.190 \\
    
    \bottomrule
\end{tabular}
\label{table:res_supervise}
\end{adjustbox}
\end{table}

\section{Conclusion}
This paper integrates the self-supervised training strategy into the domain adaptation system. Compared to the traditional UDA method, the proposed SSDA method fully leverages the potential label information from unlabeled target domain data and speaker discrimination ability from the source domain. Our proposed SSDA system achieves 10.2\% EER on the CnCeleb dataset without using any target domain speaker label and even exceeds the state-of-the-art result. Besides, when source and target domain labels are both used and are jointly trained, our system makes further improvement and achieves 8.86\% EER on the CnCeleb dataset.

\section{Acknowledgements}
This work was supported by the China NSFC projects (No. 62071288 and No. U1736202). Experiments have been carried out on the PI supercomputers at Shanghai Jiao Tong University. The author Zhengyang Chen is supported by Wu Wen Jun Honorary Doctoral Scholarship, AI Institute, Shanghai Jiao Tong University.

\bibliographystyle{IEEEbib}
\bibliography{strings,refs}

\end{document}